\documentclass[aps,prl,twocolumn,floatfix,superscriptaddress,prl]{revtex4-1}
\usepackage[latin9]{inputenc}
\usepackage{amsmath}
\usepackage{amssymb}
\usepackage{graphicx}
\usepackage{esint}
\usepackage{epstopdf}
\usepackage{braket}
\usepackage{color}
\usepackage{hyperref}
\usepackage{breakurl}
\usepackage{tikz-feynman}
\usepackage{bibunits}
\usepackage{mathtools}
\usepackage{stackengine}
\usepackage{dsfont}
\usepackage{soul}

\makeatletter
\@ifundefined{textcolor}{}
{
 \definecolor{BLACK}{gray}{0}
 \definecolor{WHITE}{gray}{1}
 \definecolor{RED}{rgb}{1,0,0}
 \definecolor{GREEN}{rgb}{0,1,0}
 \definecolor{BLUE}{rgb}{0,0,1}
 \definecolor{CYAN}{cmyk}{1,0,0,0}
 \definecolor{MAGENTA}{cmyk}{0,1,0,0}
 \definecolor{YELLOW}{cmyk}{0,0,1,0}
}

\makeatother

\begin{document}

\title{Quantum discord and steering in top quarks at the LHC}

\author{Yoav Afik}
\email{yoavafik@gmail.com}
\affiliation{Experimental Physics Department, CERN, 1211 Geneva, Switzerland}

\author{Juan Ram\'on Mu\~noz de Nova}
\email{jrmnova@fis.ucm.es}
\affiliation{Departamento de F\'isica de Materiales, Universidad Complutense de Madrid, E-28040 Madrid, Spain}

\begin{abstract}
Top quarks have been recently shown to be a promising system to study quantum information at the highest-energy scale available. The current lines of research mostly discuss topics such as entanglement, Bell nonlocality or quantum tomography. Here, we provide the full picture of quantum correlations in top quarks by studying also quantum discord and steering. We find that both phenomena are present at the LHC. In particular, quantum discord in a separable quantum state is expected to be detected with high-statistical significance. Interestingly, due to the singular nature of the measurement process, quantum discord can be measured following its original definition, and the steering ellipsoid can be experimentally reconstructed, both highly-demanding measurements in conventional setups. In contrast to entanglement, the asymmetric nature of quantum discord and steering can provide witnesses of $CP$-violating physics beyond the Standard Model.
\end{abstract}

\date{\today}

\maketitle

\textit{Introduction.}---The top quark is the most massive particle of the Standard Model, with a mass $m_tc^2\approx 173~\textrm{GeV}$~\cite{ParticleDataGroup:2018ovx}. This large mass is translated into a large decay width, which renders the top lifetime ($\sim 10^{-25}~\textrm{s}$) much shorter than the time scales of hadronization ($\sim 10^{-23}~\textrm{s}$) and spin decorrelation ($\sim 10^{-21}~\textrm{s}$). Such a fast decay makes the top quark unique, allowing us to reconstruct its spin from the decay products. Top quarks are produced in top-antitop ($t\bar{t}$) pairs, whose spin correlations have been widely studied in the literature~\cite{Kane1992,Bernreuther1994,Mahlon1996,Parke1996,Bernreuther1998,Bernreuther2001,Bernreuther2004,Uwer2005,Baumgart2013,Bernreuther2015}, devoting special attention to potential signatures of new physics violating the $CP$ invariance of the Standard Model~\cite{Atwood:2000tu}. The measurement of top spin correlations is already a well-established technique, achieved by the D0 and CDF Collaborations at the Tevatron~\cite{Aaltonen2010,Abazov2011ka,Abazov2015psg}, and by ATLAS and CMS at the Large Hadron Collider (LHC)~\cite{Aad2012,Chatrchyan2013,Aad2014mfk,Sirunyan2019,Aaboud2019hwz}. 

It was recently suggested that top quarks can be also used to study quantum information~\cite{Afik2021}, something of particular interest due to the genuine relativistic character (a critical feature in quantum information~\cite{Czachor1997,Gingrich2002,Peres2004,Friis2013,Giacomini2019,Ghodrati:2020vzm,Kurashvili2022}) and fundamental nature of the Standard Model, representing the highest-energy scale available, at the current frontier of known physics. This line of research has quickly inspired a number of works~\cite{Fabbrichesi2021,Severi2022,Aoude2022,Afik2022,Aguilar2022,Fabbrichesi2022}. Quantum phenomena can be also studied in other elementary particles such as neutrinos~\cite{Formaggio2016,Ming2020,Blasone2021}, $\tau$ leptons~\cite{Fabbrichesi2022,Altakach2023}, or massive gauge bosons~\cite{Barr2022,Barr2022B,Ashby2023,Aguilar2023,Aguilar2023a}. 

The current approaches focus on quantum tomography, entanglement, and Bell nonlocality. An alternative manifestation of quantumness is quantum discord~\cite{Ollivier2001}, the most basic form of quantum correlations, more general than entanglement (any entangled state shows discord but the opposite is not true). Because of its stronger robustness, discord has attracted attention for its potential role in quantum technologies~\cite{Datta2008,Lanyon2008,Streltsow2011,Auccaise2011QTAsu,Dakic2012,Madsen2012,Silva2013,Gessner2014,Adesso2014,Ma2016,Wang2019}. Moreover, in contrast to entanglement, discord is asymmetric between different subsystems.

Steering is another asymmetric form of quantum correlations, where measurements on one subsystem ``steer'' the quantum state of other one. Steering was the way in which Schr\"odinger~\cite{Schrodinger1935} conceived the Einstein-Podolsky-Rosen paradox~\cite{Einstein1935}, although the precise formulation of the concept had to wait for more than 70 years~\cite{Wiseman2007}. Steerability is a nonlocal feature of quantum mechanics that lies between entanglement and Bell nonlocality, giving rise to the following hierarchy of quantum correlations:
\begin{equation}\label{eq:Hierarchy}
\resizebox{\hsize}{!}{$\textrm{Bell~Nonlocality} \subset \textrm{Steering} \subset \textrm{Entanglement}\subset \textrm{Discord}.$}
\end{equation}
Steering also presents a number of potential applications as a quantum resource~\cite{Gallego2015,Chen2016,Zhang2019,Roy2020,Nguyen2020,Hao2022}.

Here we provide the full picture of quantum correlations in top quarks by studying quantum discord and steering, finding that both are present at the LHC. Specifically, discord in a separable state can be detected with high-statistical significance. Remarkably, quantum discord can be measured following its original definition~\cite{Ollivier2001}, in contrast with most experimental setups~\cite{Lanyon2008,Auccaise2011QTAsu,Madsen2012,Silva2013,Gessner2014,Adesso2014,Wang2019}. Furthermore, the steering ellipsoid~\cite{Jevtic2014}, a fundamental object in quantum information, can be experimentally reconstructed. Finally, due to the asymmetric nature of discord and steering, we show that both magnitudes can reveal signatures of new physics.

\textit{Two-qubit discord and steering.}---The most simple system displaying quantum correlations is a pair of qubits $A,B$ (Alice and Bob), described by a density matrix 
\begin{equation}\label{eq:GeneralBipartiteStateRotations}
\rho=\frac{\mathds{1}+\sum_{i}\left(B^{+}_i\sigma^i\otimes \mathds{1}+B^{-}_i \mathds{1}\otimes\sigma^i \right)+\sum_{i,j}C_{ij}\sigma^{i}\otimes\sigma^{j}}{4}
\end{equation}
with $\sigma^i$, $i=1,2,3$, the usual Pauli matrices, $\mathbf{B}^{\pm}$ the Bloch vectors of Alice and Bob, and $\mathbf{C}$ their correlation matrix. The expression for the quantum discord of Alice reads~\cite{Lu2011,Girolami2011}
\begin{equation}\label{eq:DiscordQubit}
    \mathcal{D}_{A}= S(\rho_B)-S(\rho)+\min_{\mathbf{\hat{n}}}p_{\mathbf{\hat{n}}}S(\rho_\mathbf{\hat{n}})+p_{-\mathbf{\hat{n}}}S(\rho_{-\mathbf{\hat{n}}})~,
\end{equation}
$\rho_{A,B}=\textrm{Tr}_{B,A}\rho$ being the reduced quantum states in $A,B$ and $S(\rho)=-\textrm{Tr}\rho\log_2\rho$ the Von Neumann entropy. The conditional quantum state $\rho_{\pm \mathbf{\hat{n}}}$ describes Alice's qubit after measuring Bob's qubit in the state $\ket{\pm \mathbf{\hat{n}}}$ (defined by  $\mathbf{\sigma}\cdot\mathbf{\hat{n}}\ket{\pm \mathbf{\hat{n}}}=\pm\ket{\pm \mathbf{\hat{n}}}$) with a probability $p_{\pm \mathbf{\hat{n}}}$. Specifically,
\begin{equation}\label{eq:conditionalstates2qubits}
    \rho_\mathbf{\hat{n}}= \frac{1+\mathbf{B}^+_{\mathbf{\hat{n}}}\cdot\mathbf{\sigma}}{2},~\mathbf{B}^+_{\mathbf{\hat{n}}}=\frac{\mathbf{B}^+ +\mathbf{C}\cdot\mathbf{\hat{n}}}{1+\mathbf{\hat{n}}\cdot\mathbf{B}^-},~p_\mathbf{\hat{n}}=\frac{1+\mathbf{\hat{n}}\cdot\mathbf{B}^-}{2}
\end{equation}
The classical version of Eq.~(\ref{eq:DiscordQubit}), involving probability distributions, vanishes identically as it involves the difference between two equivalent expressions for the mutual information. The minimization is performed over the surface of the Bloch sphere of Bob and describes the choice of the least disturbing measurement, quantifying in this way the actual degree of quantumness~\cite{Ollivier2001}. Bob's discord $\mathcal{D}_B$ is readily evaluated by interchanging $A\leftrightarrow B$, $\mathbf{B}^+\leftrightarrow \mathbf{B}^-$, $\mathbf{C}\leftrightarrow \mathbf{C}^{\mathrm{T}}$ in the above expressions.

The possible values of the Bloch vector $\mathbf{B}^{\pm}_{\mathbf{\hat{n}}}$ of the conditional state $\rho_{\mathbf{\hat{n}}}$ generate the so-called steering ellipsoid of Alice (Bob), which describes the set of quantum states that Bob (Alice) can steer to with local measurements~\cite{Jevtic2014}. This ellipsoid is not only relevant for steering, but also captures an important amount of information about the system~\cite{Jevtic2014}. Geometrical approaches
are in general useful tools to characterize quantum states~\cite{Horodecki1996,Dakic2012,Shi2012,Yu2018,Du2021,Seiler2021,Kumar2022}. 

\begin{figure}[tb!]\includegraphics[width=\columnwidth]{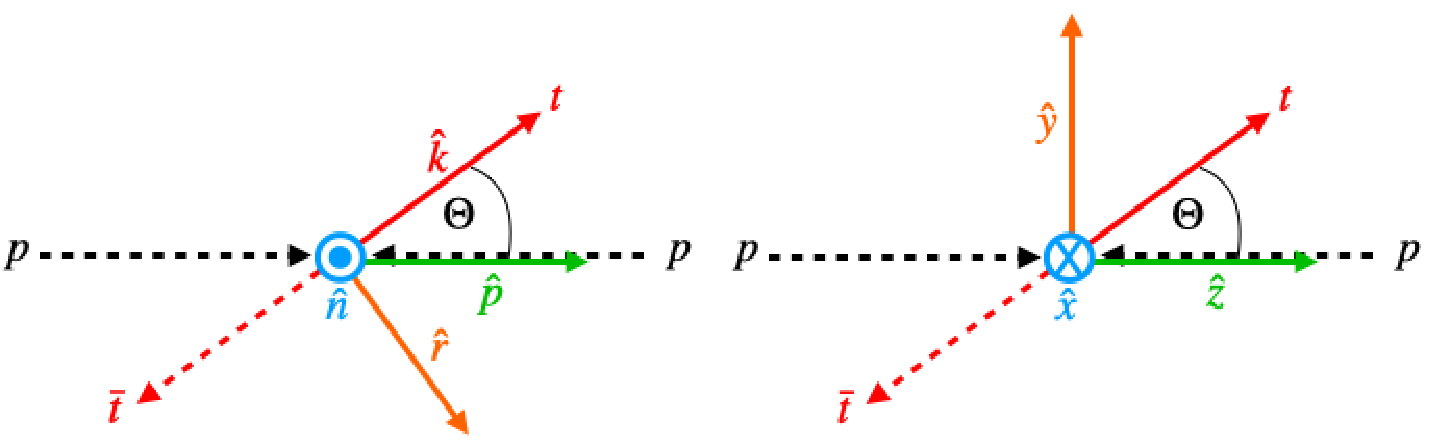}
     \caption{Orthonormal basis defined in the c.m. frame. Left: Helicity basis. Right: Beam basis.}
     \label{fig:TotalBasis}
\end{figure}

\begin{figure}[tb!]
\setlength\tabcolsep{0 pt}
\begin{tabular}{@{}ll@{}}
    \includegraphics[width=0.53\columnwidth]{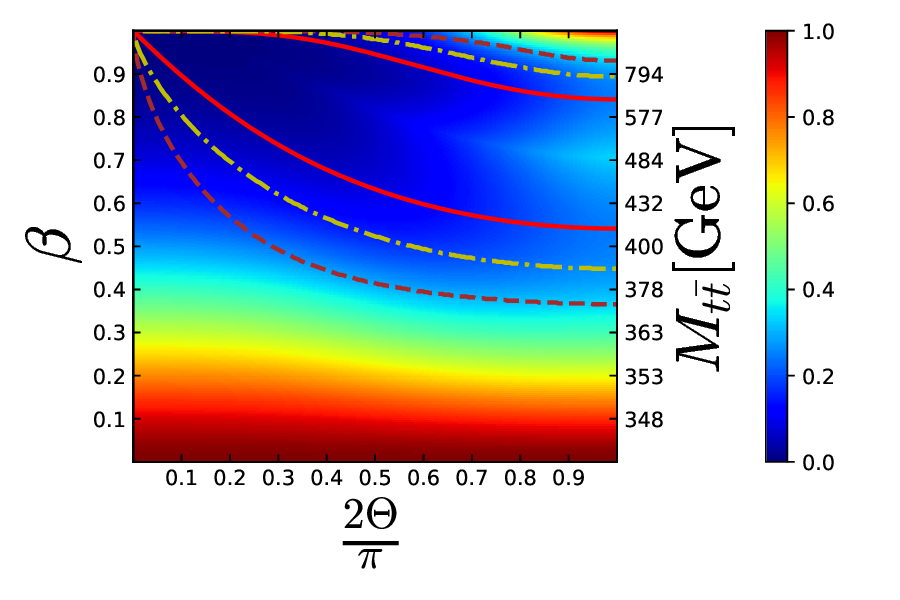} &
    \includegraphics[width=0.53\columnwidth]{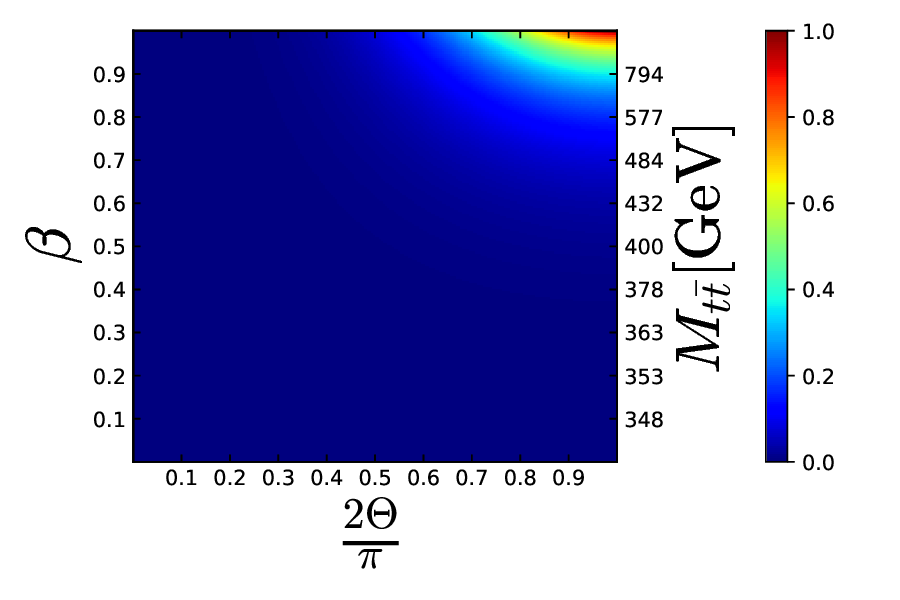} \\
     \includegraphics[width=0.53\columnwidth]{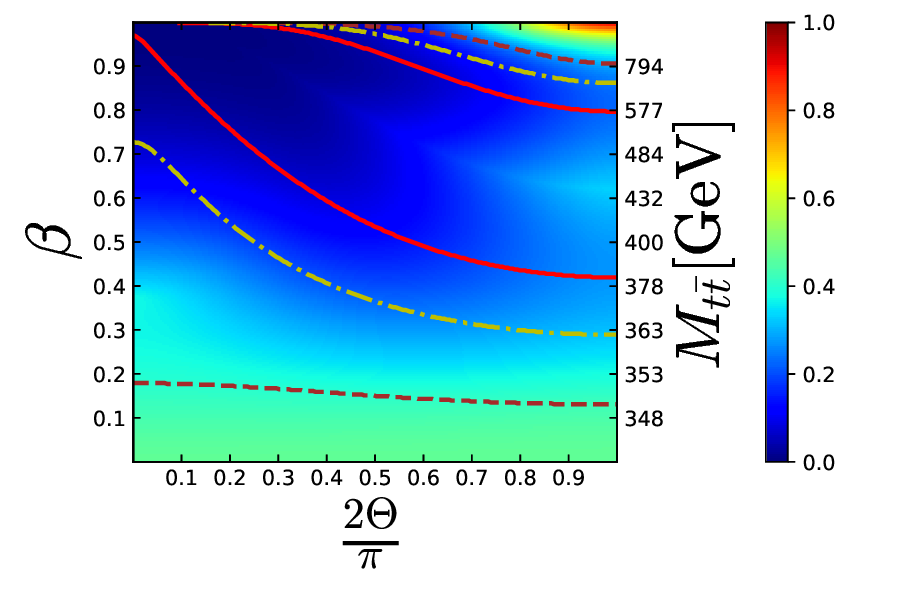} &
    \includegraphics[width=0.53\columnwidth]{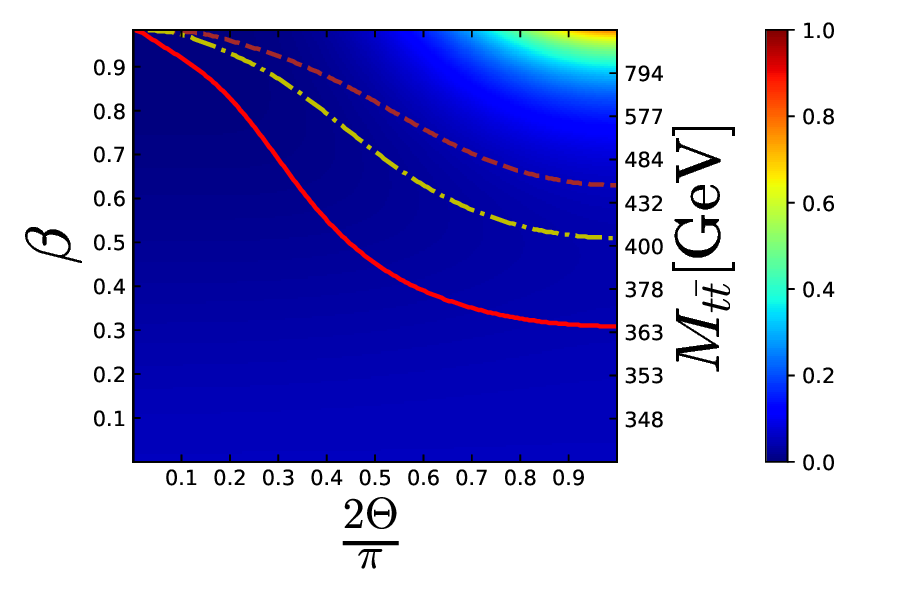}
\end{tabular}
     \caption{Quantum discord of the spin density matrix $\rho(M_{t\bar{t}},\hat{k})$ as a function of the top velocity $\beta$ and the production angle $\Theta$ in the $t\bar{t}$ c.m. frame. All plots are symmetric under $\Theta\rightarrow \pi-\Theta$. Upper left: $gg\rightarrow t\bar{t}$. Upper right: $q\bar{q}\rightarrow t\bar{t}$. Lower left: $t\bar{t}$ production at the LHC for Run 2 c.m. energy, $\sqrt{s}=13~\textrm{TeV}$~\cite{Sirunyan2019}. Lower right: $t\bar{t}$ production at the Tevatron for $\sqrt{s}=2~\textrm{TeV}$, close to its last-run c.m. energy $\sqrt{s}=1.96~\textrm{TeV}$~\cite{Abazov2015psg}. Solid red, dashed-dotted yellow, and dashed brown lines are the critical boundaries of separability, steerability, and Bell locality, respectively.}
     \label{fig:DiscordIndividual2D}
\end{figure}

\textit{Top-antitop quantum state.}---A high-energy example of two-qubit state is the spin-quantum state of a $t\bar{t}$ pair, produced from proton-proton ($pp$) or proton-antiproton ($p\bar{p}$) collisions in high-energy colliders. The $t\bar{t}$ kinematics is determined in its center-of-mass (c.m.) frame by the invariant mass $M_{t\bar{t}}$, related to the top c.m. velocity $\beta$ by $\beta=\sqrt{1-4m_t^2/M_{t\bar{t}}^2}$, and the top direction $\hat{k}$. 

The $t\bar{t}$ spin-quantum state for fixed energy and direction is described by the density matrix $\rho(M_{t\bar{t}},\hat{k})$, computed here through leading-order (LO) QCD perturbation theory since it provides simple and accurate results~\cite{Bernreuther2004,Uwer2005,Baumgart2013,Afik2021,Fabbrichesi2021,Severi2022,Aoude2022,Afik2022}. The $t$,$\bar{t}$ spins are evaluated in their respective rest frames, where they are well defined as $\rho(M_{t\bar{t}},\hat{k})$ has fixed momentum~\cite{Czachor1997,Gingrich2002,Peres2004}. Typical orthonormal basis used to characterize the vectors $\mathbf{B}^{\pm}$ and the matrix $\mathbf{C}$ are the helicity ($\{\hat{k},\hat{n},\hat{r}\}$, with $\hat{n}$ perpendicular to the scattering plane) and beam ($\{\hat{x},\hat{y},\hat{z}\}$, with $\hat{z}=\hat{p}$ along the initial beam) basis, depicted in left and right Fig.~\ref{fig:TotalBasis}, respectively. A comprehensive introduction to $t\bar{t}$ physics through a quantum information approach is presented in Ref.~\cite{Afik2022}, including the formalism behind this work. 

The approximate $CP$ invariance of the Standard Model imposes in general $\mathbf{B}^+=\mathbf{B}^-$ and $\mathbf{C}=\mathbf{C}^{\textrm{T}}$, so quantum discord and steerability are symmetric, $\mathcal{D}=\mathcal{D}_t=\mathcal{D}_{\bar{t}}$, where hereafter we identify $A,B$ with the $t,\bar{t}$ spins, respectively. Moreover, at LO, spin polarizations vanish $\mathbf{B}^{\pm}=0$, simplifying the calculations as then $\rho(M_{t\bar{t}},\hat{k})$ is a $T$ state~\cite{Horodecki1996}. Specifically, quantum discord can be computed analytically, while a sufficient and necessary condition for steerability is~\cite{Jevtic2015}
\begin{equation}\label{eq:steerability}
\int\mathrm{d}\mathbf{\hat{n}}~\sqrt{\mathbf{\hat{n}}^{\textrm{T}}\mathbf{C}^{\textrm{T}}\mathbf{C}
    \mathbf{\hat{n}}}>2\pi
\end{equation}
where the integral runs over the unit sphere.

Figure \ref{fig:DiscordIndividual2D} shows the discord of $\rho(M_{t\bar{t}},\hat{k})$, a function solely of $\beta$ and the production angle $\cos\Theta=\hat{k}\cdot \hat{p}$. In the upper row, we consider $t\bar{t}$ production from the most elementary QCD processes: an initial state of gluon-gluon ($gg$) or quark-antiquark ($q\bar{q}$). 

\begin{figure*}[tb!]
\begin{tabular}{@{}cc@{}}
    \includegraphics[width=\columnwidth]{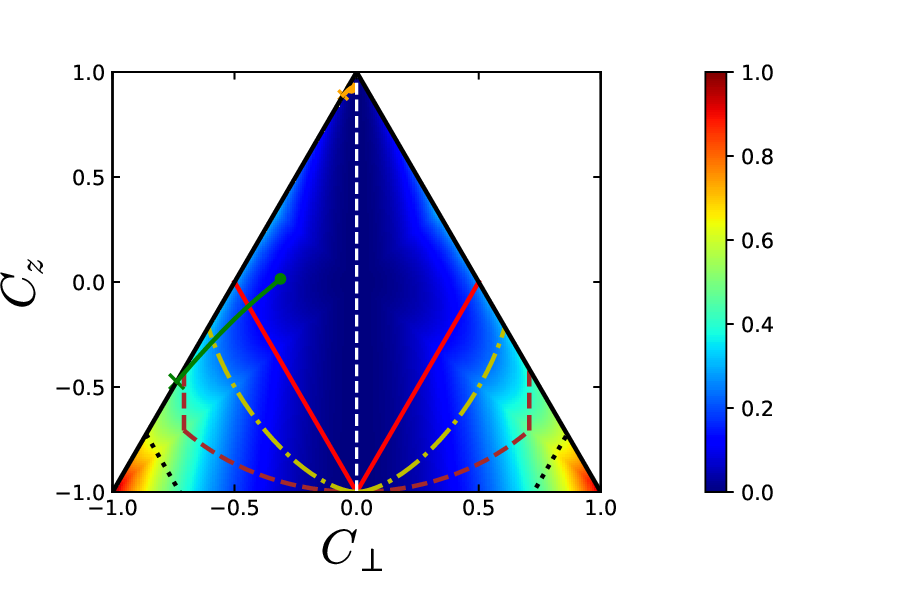} &
    \includegraphics[width=1.06\columnwidth]{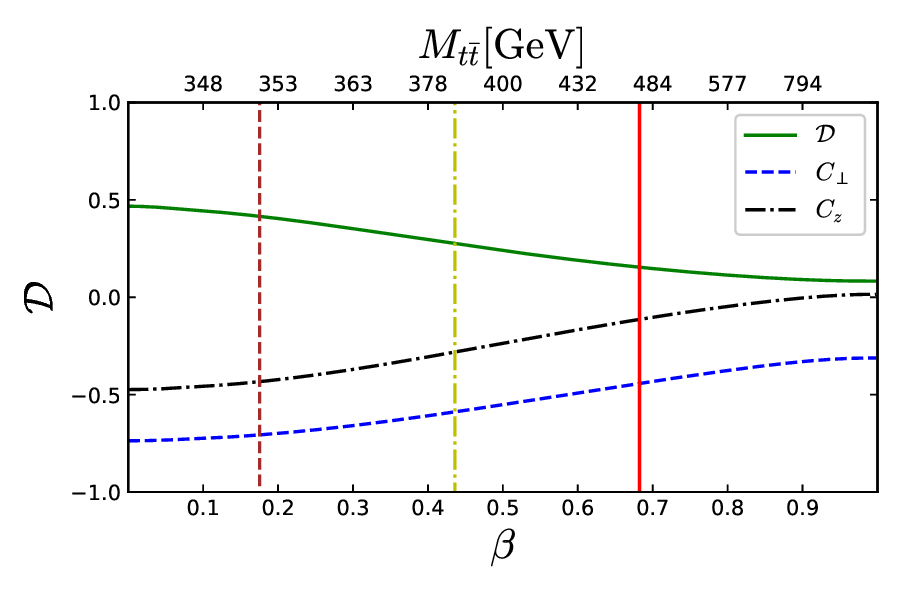}
\end{tabular}
     \caption{Quantum discord of the integrated quantum state $\rho(M_{t\bar{t}})$. Left: Discord as a general function of $C_\perp,C_z$. The colored region inside the triangle represent the physical quantum states. Vertical dashed white line marks the classical states, $C_\perp=0$. Solid red, dashed-dotted yellow, dashed brown, and dotted black lines are the critical boundaries of separability, steerability, Bell locality, and NAQC, respectively. Solid green (orange) line is the $[C_\perp(\beta),C_z(\beta)]$ trajectory for the LHC (Tevatron) at $\sqrt{s}=13~\textrm{TeV}$ ($\sqrt{s}=2~\textrm{TeV}$). The cross (circle) represent the values for $\beta=0$ ($\beta=1$). Right: Detailed trajectory of green line in left panel.} 
     \label{fig:Discord21D}
\end{figure*}

For $gg$ channel (upper left panel), discord is strong both at threshold ($\beta=0$) and for high transverse momentum $p_T$ ($\beta\to 1$ and $\Theta\to\pi/2$), where the $t\bar{t}$ pair is in maximally entangled singlet and triplet pure states, respectively. Separable states also exhibit discord, as shown by the entanglement boundaries~\cite{Afik2021} (solid red lines). The boundaries for steerability [computed from Eq.~(\ref{eq:steerability})] and Bell nonlocality~\cite{Afik2022} are given by dashed-dotted yellow and dashed brown lines. As expected, the plot follows the hierarchy (\ref{eq:Hierarchy}).

On the other hand, for $q\bar{q}$ channel (upper right panel), since $\rho(M_{t\bar{t}},\hat{k})$ is Bell nonlocal in whole phase space~\cite{Afik2022}, it is also steerable and shows discord, with $\mathcal{D}=1-h\left(1/(2-\beta^2\sin^2\Theta)\right)$, $h(x)\equiv -x\log_2x-(1-x)\log_2(1-x)$. Discord only vanishes at threshold or for forward production ($\Theta=0$) because there $\rho(M_{t\bar{t}},\hat{k})$ is a classically correlated state~\cite{Oppenheim2002} along the beam axis $\hat{p}$, $C_{ij}=\hat{p}_i\hat{p}_j$. The quantumness of $\rho(M_{t\bar{t}},\hat{k})$ only becomes appreciable for high $p_T$, where it converges to the same triplet state as $gg$ channel~\cite{Afik2021}.

The quantum state of any $t\bar{t}$ produced through QCD can be written as a convex sum of these elementary quantum states~\cite{Afik2022}. Lower left and right Fig.~\ref{fig:DiscordIndividual2D} show $t\bar{t}$ production at the LHC ($pp$ collisions) and Tevatron ($p\bar{p}$ collisions), where $gg$ and $q\bar{q}$ mechanisms dominate, respectively, as seen from the plots. Thus, Fig.~\ref{fig:DiscordIndividual2D} describes the full hierarchy of quantum correlations in $t\bar{t}$ QCD production within the Standard Model.

\textit{Integrated $t\bar{t}$ quantum state.}---In experiments, all magnitudes are integrated over phase space. For the analysis, we consider the integrated two-qubit quantum state~\cite{Afik2022}
\begin{eqnarray}\label{eq:QuantumStateTotaltt}
\rho(M_{t\bar{t}})=\frac{1}{\sigma(M_{t\bar{t}})}\int\limits^{M_{t\bar{t}}}_{2m_t}\mathrm{d}M\int\mathrm{d}\Omega~\frac{\mathrm{d}\sigma}{\mathrm{d}\Omega\mathrm{d}M} \rho(M,\hat{k}),
\end{eqnarray}
with $\frac{\mathrm{d}\sigma}{\mathrm{d}\Omega\mathrm{dM_{t\bar{t}}}}$ the differential cross section, proportional to the probability of producing a $t\bar{t}$ pair with $(M_{t\bar{t}},\hat{k})$, and $\sigma(M_{t\bar{t}})$ the integrated cross section ensuring normalization, $\textrm{Tr}\,\rho(M_{t\bar{t}})=1$. The integration limits mean that we average over all possible top directions, but only select events in the energy window $[2m_t,M_{t\bar{t}}]$. This average induces invariance under rotations around the beam axis, further imposing certain symmetries on $\rho(M_{t\bar{t}})$. Specifically, in the beam basis, the polarizations are longitudinal, $B^{\pm}_{i}=B^{\pm}_{z}\delta_{i3}$, and the correlation matrix is diagonal, $C_{ij}=\delta_{ij}C_j$, with $C_{1}=C_{2}=C_{\perp}$ and $C_{3}=C_{z}$. Moreover, at LO, $\rho(M_{t\bar{t}})$ is unpolarized so it is a simple $T$ state characterized by $2$ parameters, $C_\perp,C_z$.

Left Fig.~\ref{fig:Discord21D} shows the general dependence of $\rho$ on $C_\perp,C_z$. The colored triangle represents the physical values $1-C_{z}-2|C_{\perp}|\geq 0$ where $\rho$ is non-negative. The internal lower left and right triangles are the entanglement regions, delimited by $-C_{z}+2|C_{\perp}|=1$ (solid red). In particular, the leftmost (rightmost) vertex is a spin singlet (triplet). Once more, separable states exhibit discord. Indeed, $\rho$ is a classical state \textit{iff} $C_\perp=0$ (vertical dashed white). The steerability boundaries (dashed-dotted yellow) are 
\begin{equation}
    |C_z|+\dfrac{|C_\perp|}{\sqrt{1-\dfrac{C^2_z}{C^2_\perp}}}\arcsin \sqrt{1-\frac{C^2_z}{C^2_\perp}}=1
\end{equation}
while those of Bell nonlocality (dashed brown) read $\max(C^2_\perp+C^2_z,2C^2_\perp)=1$. For completeness, we discuss the presence of nonlocal advantage of quantum coherence (NAQC)~\cite{Mondal2017}, in which Bob (Alice) can steer the coherence of Alice's (Bob's) qubit. The NAQC occupies the highest place in the hierarchy of quantum correlations, overtaking even Bell nonlocality, and its boundaries are $|C_{z}|+2|C_{\perp}|=\sqrt{6}$ (dotted black). 

The green and orange solid lines are the trajectories $[C_\perp(\beta),C_z(\beta)]$ from $\beta=0$ (cross) to $\beta=1$ (dot) for the LHC and the Tevatron, respectively. The Tevatron curve is compressed close to classical states with $C_\perp=0,~C_z=1$, far away from the entangled regions and presenting little discord. Joining this to the relatively large experimental uncertainties expected there~\cite{Abazov2015psg}, we find unlikely even an observation of discord at the Tevatron. On the other hand, a dominant signal is predicted at the LHC, as seen from right Fig.~\ref{fig:Discord21D}. Since discord is equivalent to $C_\perp\neq 0$, after comparing with typical experimental uncertainties~\cite{Sirunyan2019} we can expect a significant observation of quantum discord in a separable state through an inclusive (i.e., with no cuts in phase space; $\beta=1$ in right Fig.~\ref{fig:Discord21D}) measurement, potentially with more than $5\sigma$. Regarding steering, the statistical significance of its observation lies between those of entanglement and Bell-nonlocality~\cite{Afik2021,Severi2022,Aguilar2022}, and can be further increased by rejecting events from the $q\bar{q}$ channel~\cite{Aguilar2022}. Finally, precisely because of its restrictive character, NAQC is not expected to be achievable in $t\bar{t}$ production at the LHC, at least within the present scheme. Nevertheless, a full dedicated analysis of the significance of all these measurements is beyond the scope of the work.

Higher-order corrections to LO amount to slightly modify the trajectory of the green and orange curves in Fig.~\ref{fig:Discord21D}, while the effect of non-vanishing polarizations is still negligible as $\mathbf{B}^{\pm}\sim 10^{-2}-10^{-3}$~\cite{Bernreuther2015,Sirunyan2019}, leaving the main results unchanged.

\textit{Experimental considerations.}---The $t\bar{t}$ quickly decays after its production in high-energy colliders. A particularly interesting final state is a dileptonic decay, where a lepton-antilepton ($\ell^-\ell^+$) pair is produced. This is because their angular distribution $p(\hat{\mathbf{\ell}}_{+},\hat{\mathbf{\ell}}_{-})$, obtained from the differential cross section of the decay, takes the simple form~\cite{Bernreuther2004} 
\begin{eqnarray}\label{eq:LeptonicCrossSectionNormalized}
p(\hat{\mathbf{\ell}}_{+},\hat{\mathbf{\ell}}_{-})=\frac{1+\mathbf{B}^{+}\cdot\hat{\mathbf{\ell}}_{+}-\mathbf{B}^{-}\cdot\hat{\mathbf{\ell}}_{-}
-\hat{\mathbf{\ell}}_{+}\cdot \mathbf{C} \cdot\hat{\mathbf{\ell}}_{-}}{(4\pi)^2}
\end{eqnarray}
with $\hat{\mathbf{\ell}}_{\pm}$ the antilepton (lepton) directions in the top (antitop) rest frames, respectively. The vectors $\mathbf{B}^{\pm}$ and the matrix $\mathbf{C}$ are precisely the $t\bar{t}$ spin polarizations and spin-correlation matrix, integrated over a selected region of phase space [as in Eq. (\ref{eq:QuantumStateTotaltt})]. Hence, instead of measuring the individual $t$,$\bar{t}$ spins on an event by event basis, their expectation values $\mathbf{B}^{\pm},~\mathbf{C}$ are directly retrieved by an appropriate fit of Eq. (\ref{eq:LeptonicCrossSectionNormalized})~\cite{Bernreuther1998,Baumgart2013,Sirunyan2019,Afik2022}. With them, one performs the quantum tomography of $\rho(M_{t\bar{t}})$~\cite{Afik2021} and evaluates both discord and steering.

Remarkably, top quarks offer the complementary possibility of reconstructing the remaining one-qubit quantum states involved in Eq.~(\ref{eq:DiscordQubit}), allowing the direct evaluation of discord following its original definition~\cite{Ollivier2001}. Specifically, the Bloch vectors $\mathbf{B}^{\pm}$ and $\mathbf{B}^{\pm}_{\mathbf{\hat{n}}}$ are obtained from the reduced angular distributions
\begin{align}
    &p(\hat{\mathbf{\ell}}_{\pm})=\int\mathrm{d}\Omega_{\mp}~p(\hat{\mathbf{\ell}}_{+},\hat{\mathbf{\ell}}_{-})=\frac{1\pm \mathbf{B}^{\pm}\cdot\hat{\mathbf{\ell}}_{\pm}}{4\pi}\\
    \nonumber &p(\hat{\mathbf{\ell}}_{\pm}|\hat{\mathbf{\ell}}_{\mp}=\mp\mathbf{\hat{n}})=\frac{p(\hat{\mathbf{\ell}}_{\pm},\hat{\mathbf{\ell}}_{\mp}=\mp\mathbf{\hat{n}})}{p(\hat{\mathbf{\ell}}_{\mp}=\mp\mathbf{\hat{n}})}=\frac{1\pm \mathbf{B}^+_{\mathbf{\hat{n}}}\cdot\hat{\mathbf{\ell}}_{\pm}}{4\pi}
\end{align}
and thus $\rho_{A,B},p_{\pm\mathbf{\hat{n}}},\rho_{\pm\mathbf{\hat{n}}}$ can be independently measured without invoking the quantum tomography of $\rho$. The actual discord is obtained by minimization over $\mathbf{\hat{n}}$.

The measurement of $\mathbf{B}^{\pm}_{\mathbf{\hat{n}}}$ also allows to experimentally reconstruct the $t,\bar{t}$ steering ellipsoid by sweeping $\mathbf{\hat{n}}$ over the Bloch sphere. We note that both experiments represent a formidable task in standard laboratory setups as they require spin measurements with enough statistics over a large number of directions. Indeed, to the best of our knowledge, no such measurement of quantum discord has been ever performed, with most measurement schemes either using quantum tomographic methods or simplified discord criteria~\cite{Lanyon2008,Auccaise2011QTAsu,Madsen2012,Silva2013,Gessner2014,Adesso2014,Wang2019}. A steering ellipsoid has only been recently observed after sampling $1000$ points of Alice's Bloch sphere, with each steered state reconstructed after $5\times 10^4$ detection events~\cite{Zhang2019}. At the LHC, this intensive sampling is automatically implemented by the continuous event recording over the span of years of each LHC Run. For example, during Run~2, $\sim 116 \times 10^6$ $t\bar{t}$ events were generated, out of them $\sim 5 \times 10^6$ are of a dileptonic decay. An even higher amount of statistics is expected for Run~3~\cite{Fartoukh:2790409}. 

\textit{New physics witnesses.}---Top spin correlations are also studied due to their particular sensitivity to new physics violating $CP$ invariance~\cite{Atwood:2000tu}, which induces asymmetries in the $t\bar{t}$ spin-quantum state. Quantum discord and steering are sensitive to those asymmetries, forbidden within the Standard Model, and thus are natural candidates to signal new physics. For instance, any nonzero value of $\Delta \mathcal{D}_{t\bar{t}}\equiv \mathcal{D}_{t}-\mathcal{D}_{\bar{t}}$ is a signature of new physics. Alternative signatures are asymmetries between the $t,\bar{t}$ steering ellipsoids, such as differences between the ellipsoid centers and/or the orientations and lengths of the semiaxes. 

In the beam basis, these asymmetries only probe nonzero values of $B^{+}_z-B^{-}_z$ as $\mathbf{C}$ is diagonal. Nevertheless, integrating the matrix elements in the helicity basis in Eq.~(\ref{eq:QuantumStateTotaltt}) yields a density matrix $\bar{\rho}(M_{t\bar{t}})$ that, although it describes a \textit{fictitious} quantum state~\cite{Afik2022}, is still a physical non-negative density matrix, and sensitive to nonzero values of $\mathbf{B}^+-\mathbf{B}^-,\mathbf{C}-\mathbf{C}^{\mathrm{T}}$~\cite{Sirunyan2019}. As a result, the discord and steering asymmetries of $\bar{\rho}(M_{t\bar{t}})$ can directly probe these $CP$-odd magnitudes.

We stress that these new physics signatures are model independent, applying regardless of the precise mechanism behind the $CP$ violation. In analogy with quantum information witnesses~\cite{Terhall2000}, they are new physics witnesses, symmetry-protected by the Standard Model, and only nonzero in the presence of new physics. This contrasts with the case of entanglement and Bell nonlocality, symmetric between the tops, where new physics effects are reduced to quantitative corrections to Standard Model contributions~\cite{Aoude2022,Fabbrichesi2022,Severi2023}.

\textit{Conclusions.}---We provide the full hierarchy of quantum correlations in top quarks by studying quantum discord and steering. Both phenomena can be potentially observed at the LHC, while their detection at the Tevatron is unlikely. Specifically, a highly significant observation of discord in a separable state through an inclusive measurement is expected at the LHC. Furthermore, the LHC offers the possibility of measuring quantum discord directly from its definition and of reconstructing the steering ellipsoid, both challenging measurements in conventional laboratory setups. Finally, the asymmetric nature of discord and steering makes them natural candidates to test $CP$-violating new physics.

From a quantum information perspective, this work pushes forward the prospect of using high-energy colliders to study quantum information. Apart from the intrinsic interest of such a fundamental environment, certain demanding measurements in conventional setups can be naturally implemented at colliders. From a high-energy perspective, this work further advocates the introduction of quantum information tools in high-energy physics. In particular, quantum discord and steering can provide new physics witnesses that extend current approaches based on entanglement~\cite{Aoude2022,Fabbrichesi2022,Severi2023}. 

Future works should perform a dedicated analysis of discord and steering measurements. An analysis of the impact of specific models of new physics in the witnesses proposed here is also an interesting extension. Another possibility is the study of quantum discord and steering in other high-energy systems~\cite{Barr2022,Barr2022B,Fabbrichesi2022,Aguilar2023a}.

\textit{Acknowledgements.}---We thank an anonymous referee of our previous work~\cite{Afik2021} for suggesting to us the possibility of studying quantum discord in top quarks. We also would like to thank M.~Lewenstein, W.~P.~Schleich, and F.~Sols for useful discussions. JRMdN acknowledges funding from European Union's Horizon 2020 research and innovation programme under the Marie Sk\l{}odowska-Curie grant agreement No 847635. 

\bibliographystyle{apsrev4-1}

\bibliography{QuantumDiscord}

\end{document}